\documentclass[prb,twocolumn,showpacs,superscriptaddress]{revtex4-1}

\usepackage{graphicx} % Include figure files
\usepackage{amssymb}
\usepackage{color}

\newcommand{\ve}[1]{\mathbf{#1}}

\begin{document}

\title{Band renormalization of a polymer physisorbed on graphene investigated by many-body perturbation theory}

\author{Peter Puschnig}\email{peter.puschnig@uni-graz.at}
\affiliation{Chair of Atomistic Modelling and Design of Materials, Montanuniversit\"at Leoben, Franz-Josef-Stra\ss e 18, 8700 Leoben, Austria}
\affiliation{Institut f\"ur Physik, Karl-Franzens-Universit\"at Graz, Universit\"atsplatz 5, 8010 Graz, Austria}

\author{Peiman Amiri}
\affiliation{Chair of Atomistic Modelling and Design of Materials, Montanuniversit\"at Leoben, Franz-Josef-Stra\ss e 18, 8700 Leoben, Austria}
\affiliation{Department of Physics, Isfahan University of Technology, 84156-83111 Isfahan, Iran}

\author{Claudia Draxl}
\affiliation{Chair of Atomistic Modelling and Design of Materials, Montanuniversit\"at Leoben, Franz-Josef-Stra\ss e 18, 8700 Leoben, Austria}
\affiliation{Institut f\"ur  Physik, Humboldt-Universit\"at zu Berlin, Newtonstraße 15, 12489 Berlin, Germany}

\date{\today}

\begin{abstract}
Many-body perturbation theory at the $G_0W_0$ level is employed to study the electronic properties of poly(\emph{para}-phenylene) (PPP) on graphene. Analysis of the charge density and the electrostatic potential shows that the polymer-surface interaction gives rise to the formation of only weak surface dipoles with no charge transfer between the polymer and the surface. In the local-density approximation (LDA) of density-functional theory,  the band structure of the combined system appears as a superposition of the eigenstates of its constituents. Consequently, the LDA band gap of PPP remains unchanged upon adsorption onto graphene. $G_0W_0$ calculations, however, renormalize the electronic levels of the weakly physisorbed polymer. Thereby, its band gap is considerably reduced compared to that of the isolated PPP chain. This effect can be understood in terms of image charges induced in the graphene layer, which allows us to explain the quasi-particle gap of PPP versus polymer-graphene distance by applying a classical image-potential model. For distances below 4.5 {\AA}, however, deviations from this simple classical model arise which we qualitatively explain by taking into account the polarizablity of the adsorbate. For a quantitative description with predictive power, however, we emphasize the need for an accurate \emph{ab-initio} description of the electronic structure for weakly coupled systems at equilibrium bonding distances.
\end{abstract}


%31.70.-f 	Effects of atomic and molecular interactions on electronic structure (see also section 34 Atomic and molecular collision processes and interactions)
%31.70.Dk 	Environmental and solvent effects
%71.10.-w 	Theories and models of many-electron systems
%71.15.Mb 	Density functional theory, local density approximation, gradient and other corrections 
%73.20.-r 	Electron states at surfaces and interfaces
%85.65.+h 	Molecular electronic devices

\maketitle

\section{Introduction}

When a molecule is brought in contact with a surface,
its electronic states are generally renormalized in terms of their energy positions
and widths, where the magnitude of these effects depends on the adsorption distance. 
Molecule-surface interactions may give rise to several phenomena
including hybridization of molecular states with substrate levels, charge transfer between the 
molecule and substrate, and formation of strong short-range interface dipoles. 
All of these effects can be described within a static mean-field approach as entailed in density-functional 
theory by solving single-particle Kohn-Sham (KS) equations and employing
standard local (LDA) or semi-local (GGA) approximations for the exchange-correlation potential.

An additional effect that can greatly alter the level positions
of the molecule, but is not accounted for in such a mean-field approach, arises due to polarization effects. 
Electron addition (removal) energies of the molecule, \emph{i.e.}, the quasi-particle
energies of the molecule, inherently involve a negatively (positively) charged molecule residing 
on top of a polarizable substrate. Thus, electro-static Coulomb forces of 
this added electron or hole polarize the underlying surface which, in turn, affects
the energy position of the added electrons or holes, respectively. In particular,
the ionization potential and electron affinity level, and hence the
band gap of the molecule, are renormalized due to the presence of the 
substrate.\cite{Neaton2006,Johnson1987} 
Even for physisorbed and weakly coupled molecules, these surface polarization effects
can considerably change the size of the gap of the molecule compared to its gas phase value. 

The response of the electronic system to the added electron/hole is not captured by local or semi-local KS functionals. In order to account for this dynamical polarization effect, we employ many-body perturbation theory on top of DFT calculations. In the $G_0W_0$ approximation, the self-energy $\Sigma$ is given as the product of the noninteracting single-particle Green function, $G_0$, and the screened Coulomb interaction, $W_0$, calculated within the random-phase approximation.\cite{Onida2002} In this approach, correlation effects including the response of the electronic system to the added electrons or holes are taken into account by the nonlocal and energy-dependent self-energy. As already demonstrated for the case of a benzene molecule physisorbed on various metallic and semi-conducting substrates,\cite{Neaton2006,Garcia-Lastra2009} the $G_0W_0$ band gap of the molecule is reduced relative to its corresponding gas-phase value by an amount that depends on the polarizability of the surface. In contrast, DFT calculations employing LDA or hybrid exchange-correlation functionals like PBE0, render the size of the benzene HOMO-LUMO gap independent of the underlying substrate.

In this paper, we explore how the electronic structure of a one-dimensional system --  the polymer poly(\emph{para}-phenylene) (PPP) used as active material in blue light-emitting
diodes \cite{Tasch97} -- is affected by the presence of the two-dimensional graphene layer. We thereby extend  previous \emph{ab initio} work\cite{Neaton2006,Garcia-Lastra2009}  which only treated small organic molecules on various surfaces. We study the renormalization of the molecular electronic levels by employing $G_0W_0$ calculations. 
%CD: Do we need this? If yes, we should cite the experiments.
%{\red While several experimental and theoretical studies on the electronic properties of free-standing graphene and on isolated PPP exist in the literature, to the best of our knowledge, the adsorption of PPP on graphene has not been investigated before. }
We first provide details on our computational methodology concerning the DFT and $G_0W_0$ calculations. Then, the electronic band structure of graphene is discussed, both in the hexagonal unit cell and in a larger rectangular supercell necessary for the calculations of the adsorbate system. The electronic band structure at the LDA and $G_0W_0$ level of isolated PPP is presented and compared to available literature results. As the main results, we discuss the combined system (denoted as PPP@gr) showing that the polymer's quasi-particle HOMO-LUMO gap is strongly renormalized due to the presence of the graphene layer. We model our \emph{ab initio} results by a classical image-potential model in order to describe the asymptotic behavior of the potential felt by an electron outside the substrate. We also extend the standard image-charge model as to include the polarizability of PPP leading to a better model-description of the $G_0W_0$ results at short polymer-graphene separations.
       
\section{Computational approach}
\label{sec:comp}
Self-consistent DFT calculations are performed by using ABINIT,\cite{Gonze2009} which is a plane-wave based code with periodic boundary conditions along all three directions. We utilize norm-conserving pseudo-potentials generated by the Troullier-Martins scheme.\cite{Troullier1993} Exchange-correlation effects are treated within the local-density approximation. An energy cutoff of 50 Ry is chosen for the electronic wave functions. Brillouin zone integrations are carried out by using Monkhorst-Pack \cite{Monkhorst1976} meshes of (5$\times$17$\times$1) $k$ points for the PPP@graphene supercell along with a Methfessel-Paxton\cite{Methfessel1989} smearing with a smearing parameter of 0.01 Ry. The so obtained DFT ground-state results serve as a starting point for the subsequent $G_0W_0$ computations.

In the ${G}_{0}{W}_{0}$@LDA method, the QP energies are obtained from the linearized QP equation:
\begin{equation}
\label{gw}
\varepsilon^{QP}_n = \varepsilon_n^{LDA} 
+ Z_n{\left<\psi_n^{LDA}\mid\Sigma(\varepsilon_n^{LDA}) - V_{xc}\mid \psi_n	^{LDA}\right>}.
\end{equation}
Here, $\psi_n^{LDA}$ and $\varepsilon_{n}^{LDA}$ are DFT-LDA eigenstates and eigenvalues, and the renormalization factor $Z_n$ is given by
\begin{equation}
\label{weight}
Z_n= {\left[1 - \left. \frac{\partial\left< \psi_n^{LDA} \mid \Sigma(\varepsilon) \mid\psi_n^{LDA}\right>}
{\partial\varepsilon} \right|_{\varepsilon_{n}^{LDA}}\right]}^{-1}.
\end{equation}
The self-energy, $\Sigma$, is calculated non-self-consistently from the convolution of the non-interacting single-particle Green function $G$ with the screened Coulomb interaction $W_0$, \emph{i.e.}, $\Sigma=iG_0W_0$. Here, the Green function is defined as $G_0(z)=(z-H^{LDA})^{-1}$ where the subscript $0$ symbolizes the fact that these quantities are obtained from the Kohn-Sham orbitals and energies in a non-self-consistent manner.
Vertex corrections are neglected both in the self-energy and in the polarizability and, hence, in the calculation of $W_0$. In this work, the frequency dependence of $W_0$ is described by a plasmon-pole model (PPM).\cite{Godby1989} 

In the calculation of the QP energies, there are two main technical problems which make the $G_0W_0$ approach computationally costly for large unit cells as necessary for studying the adsorption of molecules or polymers on surfaces. The first bottleneck arises from the summation over unoccupied states which appears in the correlation part of the self-energy. To overcome this problem, we make use of the recently developed energy effective technique (EET).\cite{Berger2010} In this method, all necessary steps in a $G_0W_0$ calculation can be restricted to occupied states only. As demonstrated in the Appendix for graphene (Fig.~\ref{sos-gap}), this approximation preserves the precision of the conventional sum-over-empty-states approach but speeds up calculations by more than an order of magnitude. 

The second difficulty originates from the periodic boundary conditions in combination with the repeated-slab approach to model surfaces. Thereby, one commonly constructs unit cells containing a vacuum layer to separate periodic images in order to avoid spurious interaction between them. In ground-state calculations, this poses no big problem since electronic wavefunctions and the (semi)local exchange-correlation potential decay rapidly into the vacuum, and a moderate thicknesses of the vacuum slab suffices. For $GW$ calculations, the main difficulty arises from the nonlocal nature of the self-energy, in particular, nonlocal Coulomb matrix elements, which lead to long-range image charge effects that converge very slowly as a function of the inserted vacuum layer.\cite{Onida95} To overcome this obstacle, we make use of a truncated Coulomb potential which prevents the interaction between periodic images. We utilize Ismail-Beigi's method\cite{Ismail-Beigi2006} for the case of sheet-like geometries with one confined and two periodic directions. The truncation length is chosen to be half the lattice parameter perpendicular to the surface (in $z$ direction). Using this method, $G_0W_0$ band energies converge fast as a function of vacuum layer thickness as can be seen from Fig.~\ref{qp-gap-gr} in the Appendix. Care must, however, been taken in the choice of the $k$-grid since 
the convergence with respect to the number of $k$-points is somewhat slowed when using
a cut-off Coulomb potential compared to the convergence behavior for the plain Coulomb potential
(which will be seen from the right panel of Fig.~\ref{qp-gap-gr}).
As in the public version of ABINIT, the truncation of the Coulomb interaction for slab geometries 
is not fully implemented, we have added one term taking the  $q \rightarrow 0$ limit. This way, 
we could assure fast convergence of results with respect to $k$-grid and vacuum size. 
 
\section{Results}

\subsection{Graphene}

\begin{figure}
\includegraphics*[width=0.7\columnwidth]{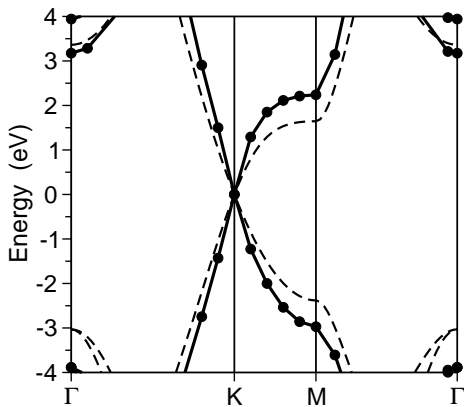}
\caption{\label{band-gw-gr}
Band structure of graphene within DFT-LDA (dashed lines)
and within the $G_0W_0$ approach (circles and full lines).
The Fermi energy is set to zero.}
\end{figure}

\begin{figure}
\includegraphics*[width=\columnwidth]{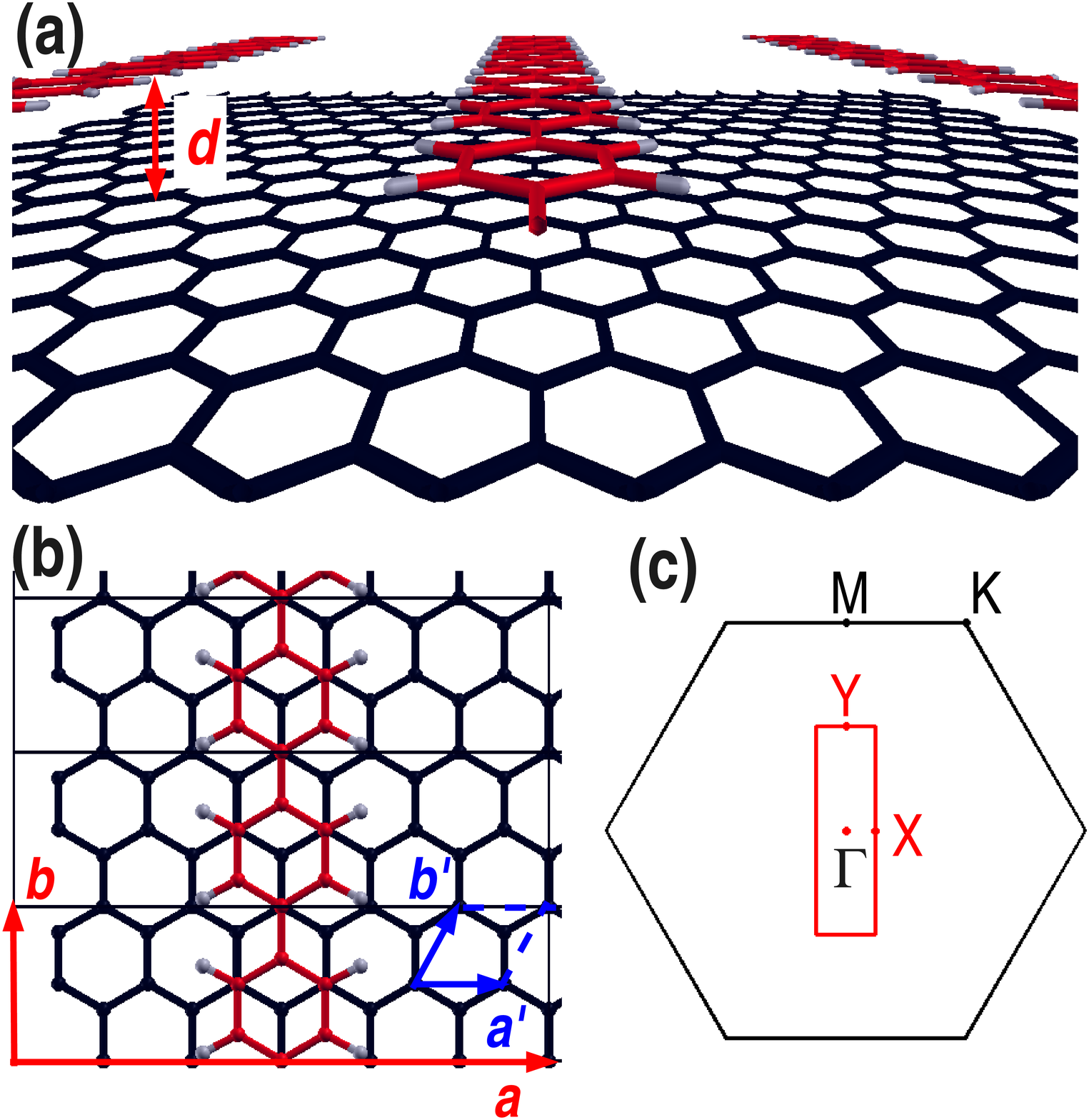}
\caption{\label{structure}
(a) Perspective side view of the structural model describing poly(\emph{para}-phenylene) (red) adsorbed on graphene (black). The adsorption height $d$ is indicated. (b) Top view showing the
unit cell vectors of the primitive, hexagonal graphene cell (blue arrows) 
as well as the rectangular supercell spanned by vectors $a$ and $b$ (red arrows).
(c) Brillouin zones corresponding to hexagonal graphene (black) as well
as to the supercell (red). High symmetry points are indicated. }
\end{figure}

Before we presents results for the adsorbed polymer on graphene (compare Fig.~\ref{structure}), we review the band structure of uncovered graphene at the LDA and $G_0W_0$ level. This is depicted in Fig.~\ref{band-gw-gr} where the LDA ($G_0W_0$) band structure is shown as dashed (solid) lines. As already noted earlier,\cite{Trevisanutto2008} the $G_0W_0$ corrections slightly enhance the overall band widths of $\pi$ and $\sigma$ bands; for instance, they increase the gap at the $\Gamma$ point from 6.6 eV to 7.0 eV. Close to the Dirac point ($K$), the Fermi velocity resulting from the LDA dispersion is about $1.01\times10^6$ ms$^{-1}$ while the $G_0W_0$ value of $1.11\times10^6$ ms$^{-1}$ is about 10\% larger and improves the agreement with experiment ($1.1\times10^6$ ms$^{-1}$).\cite{Y.Zhang05} These results emphasize the fact that
the $G_0W_0$ quasi-particle band structure not only provides improved band gaps but also leads to an improved description of graphene in terms of band widths and Fermi velocities.

In order to study the adsorption of PPP on graphene, we need to construct an appropriate supercell which is depicted in Fig.~\ref{structure}.  
If we denote the primitive lattice vectors of graphene as $\ve{a}'$ and $\ve{b}'$, depicted
as blue arrows in Fig.~\ref{structure}, the supercell vectors are given by $\ve{a}=6\ve{a}'$
and $\ve{b}=2\ve{b}'-\ve{a}'$ (red arrows) thereby spanning a rectangular supercell containing
24 carbon atoms in the graphene layer. 
When the polymer chain direction is chosen to be parallel to $\ve{b}$, its repeat unit is commensurate with the graphene basis vectors. Since we intend to study the behavior of a single PPP chain on graphene our choice of $\ve{a}=6\ve{a}'$ leads to a polymer chain separation which is large enough to prevent interaction between periodic replica of PPP chains.

\begin{figure}
\includegraphics*[width=\columnwidth]{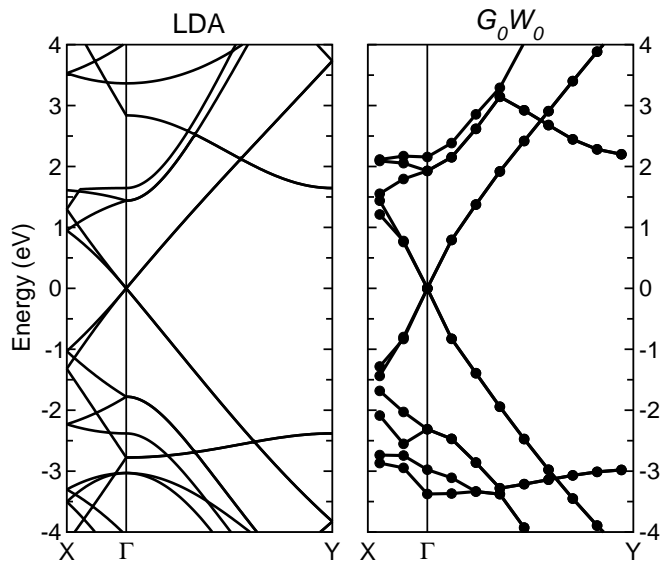}
\caption{\label{band-gw-Gr}
DFT-LDA (left) and $G_0W_0$ (right) band structure of uncovered graphene calculated in the supercell (see text). The Fermi energy is set to zero.}
\end{figure}

Fig.~\ref{band-gw-Gr} shows the band structure of \emph{uncovered} graphene in the above mentioned supercell containing 24 carbon atoms. This serves as a consistency test of our computational approach since for \emph{uncovered} graphene bands of the primitive hexagonal cell (compare Fig.~\ref{band-gw-gr}) are simply folded into the Brillouin zone corresponding to the supercell. As shown in Fig.~\ref{structure}c, the Dirac point, $K$, is folded to $\Gamma$, such that in the supercell there is a four-fold degeneracy at $\Gamma$ leading to 4 touching cones. In $\Gamma$-X direction, just below the Fermi energy, there are two bands with slightly different slopes which are arising from the original Dirac cone along the directions $K\Gamma$ and $KM$.

\subsection{Isolated PPP}

\begin{figure}
\includegraphics*[width=\columnwidth]{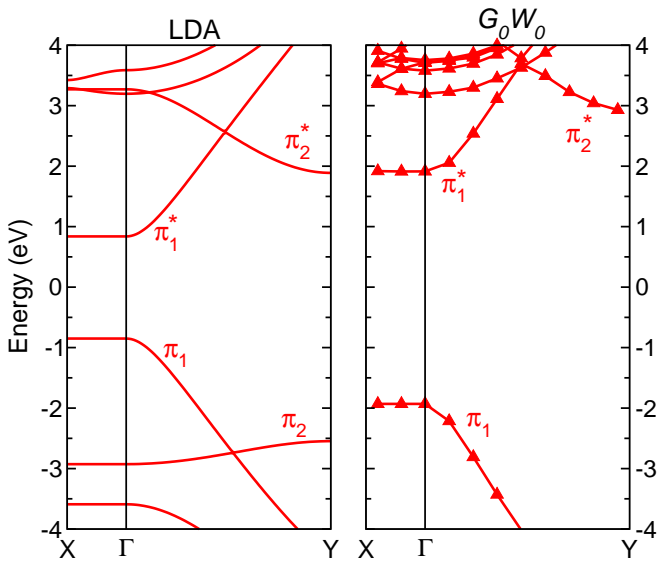}
\caption{\label{band-gw-PPP}
DFT-LDA (left) and $G_0W_0$ (right) band structure of poly(\emph{para}-phenylene) (PPP). The Fermi energy is set to zero at the mid-gap energy. The frontier $\pi$ orbitals are indicated as explained in the text.}
\end{figure} 

Before we investigate the adsorption of PPP on graphene, we present the electronic structure for an isolated PPP chain computed in the supercell introduced previously. 
The QP band structure obtained by $G_0W_0$@LDA is compared to LDA results in Fig.~\ref{band-gw-PPP}. The valence band structure of PPP close to the Fermi level is characterized by two $\pi$ bands. The inter-ring bonding band, $\pi_1$, is strongly dispersing from $\Gamma$ down to $Y$ with an LDA ($G_0W_0$) band width of
3.54 eV (3.70 eV), while the inter-ring non-bonding band, $\pi_2$, exhibits a much smaller dispersion along $\Gamma Y$. Clearly, all bands along $\Gamma X$ direction, \emph{i.e.} perpendicular to the polymer chain,  exhibit a negligible dispersion reflecting the small inter-chain interaction. The frontier unoccupied bands $\pi_1^*$ and $\pi_2^*$ are the anti-bonding counterparts of $\pi_1$ and $\pi_2$, respectively. 
The general trend of the self-energy corrections is to lower the occupied bands and raise the unoccupied bands relative to the LDA results. This leads to an increase of the direct gap at the $\Gamma$ point from 1.69 eV at the LDA level to 3.88 eV for the $G_0W_0$ band structure. This finding is in accordance with previous calculations,\cite{cad95-1,Bogar1997,Artacho2004} when considering the fact that we neglect a possible torsion angle between adjacent phenyl rings and treat PPP as a perfectly planar $\pi$-conjugated polymer.

\subsection{PPP adsorbed on graphene}

\begin{figure}
\includegraphics*[width=\columnwidth]{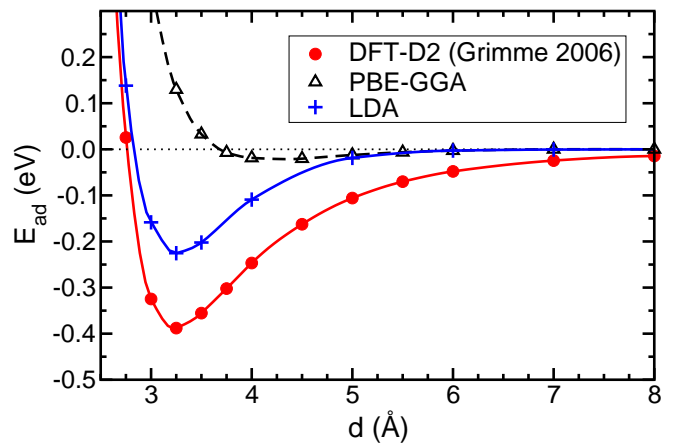}
\caption{\label{vdW}
Adsorption energy of PPP on graphene versus polymer-graphene distance $d$.
Crosses are obtained by LDA, open triangles by using the PBE-GGA functional,\cite{Perdew1996} and filled circles 
by including an empirical van-der-Waals correction according to Grimme.\cite{Grimme2006}}    
\end{figure}

We now proceed to the main outcome of the paper, \emph{i.e.} the electronic structure of
PPP adsorbed on graphene. As a prerequisite, we first investigate the adsorption geometry by
varying the polymer-graphene distance and the adsorption site in DFT total-energy calculations.
Fig.~\ref{vdW} shows the adsorption energy $E_{ad}$ as a function of adsorption height for the situation
when the center of a PPP ring is positioned on top of a graphene carbon atom, as depicted 
in Fig.~\ref{structure}b, similar to the A-B stacking in bilayer graphene.
The adsorption energy as well as the adsorption distance is very sensitive to the choice of the
exchange-correlation potential. While LDA predicts an optimal adsorption distance of about $3.25$~{\AA} with
an adsorption energy of about 0.25 eV, GGA\cite{Perdew1996} results in almost no binding (0.02 eV)
at the rather large distance of 4.5~{\AA}. This is, of course, indicative of a strong van-der-Waals
contribution to the bonding.\cite{Nabok2008} For simplicity, we employ here an empirical scheme\cite{Grimme2006} to correct
for the missing dispersion forces in GGA, resulting in an adsorption distance of $d=3.25$~{\AA} and a binding
energy of $E_{ad}=0.39$~eV as can be seen from Fig.~\ref{vdW} (red circles and line).
When comparing the A-B type of adsorption site with an A-A type of adsorption position in which
the hexagon of the polymer is on top of a graphene hexagon, we find an adsorption energy which
is by 0.10 eV less favorable. Based on the van-der-Waals corrected GGA,\cite{Grimme2006} the most favorable adsorption
site of PPP on graphene is thus analogous to the A-B stacking in bilayer graphene and also in accordance
with the adsorption position of a single benzene ring on graphite.\cite{Neaton2006}

\begin{figure}
\includegraphics*[width=0.8\columnwidth]{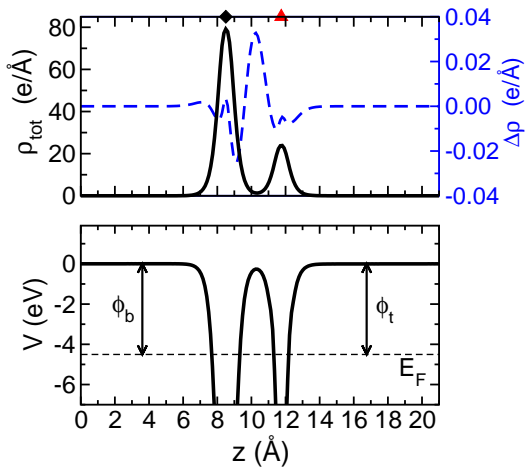}
\caption{\label{charge-potential}
Plane-averaged total charge density and charge density difference (top panel) as well as the electrostatic potential (lower panel) perpendicular to the graphene plane. The positions of the graphene plane (diamond) and the polymer chain (triangle), the Fermi level, $E_F$, and the work function on bottom and top sides of the slab $\phi_{b}$ and  $\phi_{t}$ are indicated.}    
\end{figure}

From the discussion above, it is evident that the bonding between the polymer and graphene is mainly due to van-der-Waals
interactions. This is further emphasized by analyzing the charge-density rearrangements upon adsorption. To this end,
we explore the charge density difference defined in the following way.
\begin{equation}
\label{charge-density-diff}
\Delta\rho= \rho - (\rho_{pol}+\rho_{gr}). 
\end{equation}
Here $\rho$ denotes the charge density of the combined system, while $\rho_{pol}$ and $\rho_{gr}$ are the charge densities of isolated polymer and graphene, respectively. The plane-averaged charge density difference as well as the electro-static potential is depicted in Fig.~\ref{charge-potential}. We can see that the interaction leads to regions of minor charge accumulation $(\Delta\rho>0)$  and charge depletion $(\Delta\rho<0)$ between the two constituents, \emph{i.e.}, to the formation of small surface dipoles. There is, however, no net charge transfer between the polymer and graphene. Consequently, the work-function modification due to the polymer adsorption is negligible, thus the values on the bottom and top side of the slab, $\phi_{b}$ and  $\phi_{t}$ respectively, agree with each other. These findings further prove that PPP is weakly physisorbed on graphene.
\begin{figure}
\includegraphics*[width=\columnwidth]{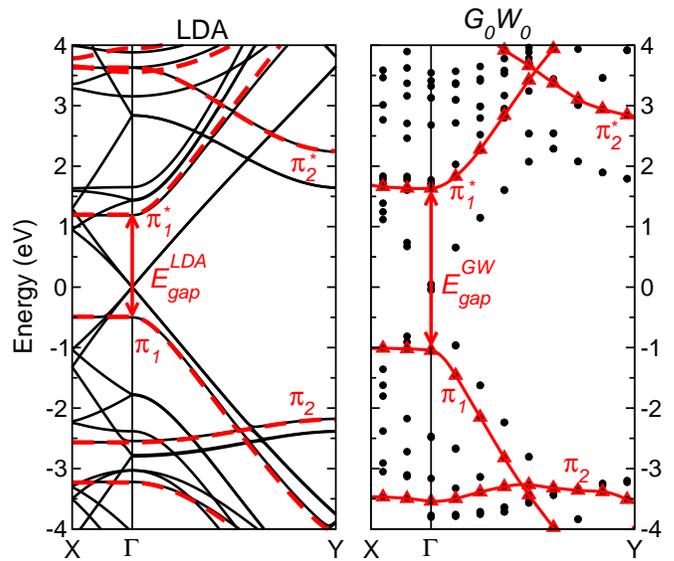}
\caption{\label{band-gw-PPP+gr}
DFT-LDA (left) and $G_0W_0$ (right) band structure of PPP adsorbed
on graphene for an adsorption height of 4 {\AA} shown as continuous black lines.
The LDA band structure of an isolated PPP chain is shown as red, dotted line and used 
to identify PPP derived states in the combinbed system. The $G_0W_0$ band structure of the frontier
PPP states is indicated by the red filled triangles, connecting lines serve as a guide for the eye.
The $G_0W_0$ enegies arising from graphene are shown as black filled circles. }
\end{figure}

In what follows, we present the band structure of the combined PPP-graphene system at the LDA as well
as at the $G_0W_0$ level. In order to highlight the effect of image charges induced in the graphene layer
on the electronic bands of PPP, we not only present data for the optimal adsorption height discussed above, but we 
also study the dependence of the band structure on the adsorption distance $d$. We start the discussion
by presenting the band structure of the adsorbate system at $d=4.0$~{\AA} as shown in Fig.~\ref{band-gw-PPP+gr}. 
Let us first focus on the LDA band structure depicted in the left panel of Fig.~\ref{band-gw-PPP+gr}. 
Here, the solid black lines depict the bands of the combined system while the red, dashed lines indicate
the bands of PPP obtained from a calculation for the isolated polymer. We observe that the LDA band
structure of the combined system appears as a mere superposition of the band strcutures of the 
isolated constituents, \emph{i.e.} graphene and PPP. The relative alignment of the two subsystems' band structures  
follows the simple rule of vacuum level alignment, also known as the Schottky-Mott limit.\cite{Schottky1938}
Thus, also the HOMO-LUMO gap of the polymer, indicated by the red arrow at the $\Gamma$ point,
remains unaltered at the level of LDA-DFT. We note that at somewhat smaller polymer-graphene distances, such as the equilibrium distance
of 3.25~{\AA}, $\pi$ wave functions of the polymer and graphene start to overlap slightly leading to a weak hybridization between the polymer and graphene states.
Correspondingly, also small charge re-arrangements and modifications of the electrostatic potential due to the equilibration of the chemical potentials of the two materials in contact
take place as indicated in Fig.~\ref{charge-potential}. However, these changes are of the order of 0.1 eV only. 
For weakly interacting systems, that are van-der-Waals bonded, we can summarize that the LDA band structure of the combined system is given by a superposition of the individual levels of the isolated subsystems, the polymer and the graphene sheet in our case, respectively. This is also the reason, why an almost substrate-independent behavior of the electronic properties of physisorbed organic molecules has been observed at the LDA-DFT level.\cite{Garcia-Lastra2009}

Our findings for the the $G_0W_0$ band structure, as depicted in the right panel of Fig.~\ref{band-gw-PPP+gr}, are in stark contrast to this substrate- and adsorption-distance-independent behavior of the LDA-DFT band structure. Here, the $G_0W_0$ band structure of the combined system is shown as black, filled circles where the
frontier PPP bands are highlighted by the red triangles and continuous lines. When compared to the $G_0W_0$ results for the isolated PPP cahin (Fig.~\ref{band-gw-PPP}),
a reduction of the HOMO-LUMO gap of PPP is evident. While the isolated chain exhibits a gap of 3.95 eV, the value of the adsorbed PPP at a distance of 4.0~{\AA} is reduced to 2.70 eV. This is a consequence of long-range correlation effects as a response of surface electrons to an added electron or hole in the polymer. This phenomenon is captured within
the $G_0W_0$ approach by the screened Coulomb potential $W_0$. For molecules on surfaces, this effect is particularly important as it contains the attractive interaction between the added electron or hole and its induced image charge. It can lead to a considerable reduction in the adsorbate's energy gap -- about 1.2 eV for the situation described above, where the band gap renormalization depends (i) on the polarizability of the surface and (ii) on the adsorbate's distance from the surface. Both effects are absent in the DFT electronic structure at the LDA or GGA level due to the locality of exchange-correlation potential, and this deficiency can also not  be cured by using hybrid functionals.\cite{Garcia-Lastra2009}

\subsection{Adsorption distance dependence}

\begin{figure}
\includegraphics*[width=\columnwidth]{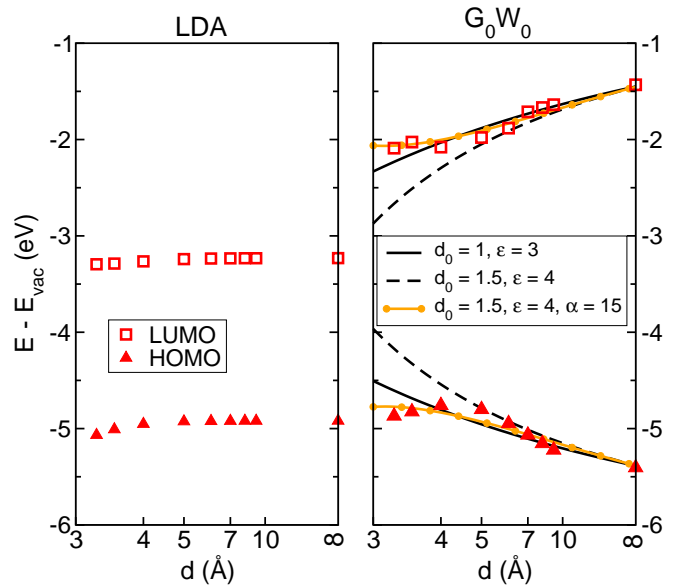}
\caption{\label{classical-model-f}
The energy of the HOMO (filled triangles) and LUMO (open squares) of PPP as a function of the adsorption distance $d$ on a reciprocal scale. The left panel shows LDA results while the right panel displays corresponding $G_0W_0$ values. The lines represent various electro-static models as described in the text. Note that energies are measured with respect to the vacuum level.}
\end{figure} 

To further emphasize the physical origin of the band-gap reduction upon adsorption, we study its dependence on the adsorption distance $d$. The results are depicted in Fig.~\ref{classical-model-f} in which we plot the energies of the HOMO and LUMO of PPP at the $\Gamma$ point as a function of PPP-graphene distance $d$, where
the left (right) panel displays LDA ($G_0W_0$) results. Note that the limiting values for $d=\infty$ are taken from calculations for an isolated PPP chain. Let us first discuss the distance dependence for moderately large values of $d \gtrsim 5$~\AA. Here, the HOMO and LUMO energies, \emph{i.e.}, the ionization and electron affinities levels of PPP, are independent of $d$ in LDA while they show strong and opposite trends in the $G_0W_0$ calculations leading to the above mentioned reduction of the band gap. This asymptotic distance-dependence can be understood by employing a classical image-potential model of the form\cite{White1998}
\begin{equation} 
\label{eq:image1}
V_{im}=-\frac{1}{4} \frac{q q'}{d-d_0}.
\end{equation}
Here, $q'=-q(\epsilon-1)/(\epsilon+1)$ is the induced image charge, where $\epsilon$ is the relative dielectric constant of the polarizable medium, and $d_0$ denotes the effective position of the image plane.
It describes the electrostatic interaction between charges (added electrons or holes) above the surface and the polarization charge below the surface. This interaction also constitutes the basic term of the $G_0W_0$ self-energy, \cite{Rohlfing2003} and is consistent with the physical picture outlined above, \emph{i.e.}, that the level re-normalizations are determined by long-range Coulomb interactions which decay very slowly with distance. The solid, black lines in
Fig.~\ref{classical-model-f} display the expected image-potential corrections according to Eq.~(\ref{eq:image1}) when inserting an effective image plane position $d_0=1.0$~{\AA} and a dielectric constant of $\epsilon=3$. The former value is in accordance with the image plane position computed for benzene adsorbed on graphite,\cite{Neaton2006} while the latter agrees with a recent result by Wehling {\emph et al.}~who calculated the dielectric constant of graphene to be 2.4.\cite{Wehling2011} Using these model parameters, the image model (solid line) only roughly follows the $G_0W_0$ values (symbols). It underestimates the level re-normalizations for large distances, particularly so for the HOMO, and does not capture the fact that the $G_0W_0$ electron affinity level seems to saturate at polymer-graphene distances smaller than 4~{\AA}, and also can not explain the more complicated behavior of the ionization potential at small $d$. 

We further investigate possible reasons for this deviation. First, a better agreement between the image model (\ref{eq:image1}) and $G_0W_0$ results for \emph{large} distances can be achieved when allowing the position of the effective image plane to move further away from the graphene plane, e.g., to $d_0=1.5$~{\AA} and also assuming a slightly bigger $\epsilon$  of 4 (black, dashed lines). These numbers lie still within the range of values reported for similar surfaces in an earlier work,\cite{Garcia-Lastra2009} and improve the agreement for separations $d \gtrsim 5$~\AA, while they clearly worsen the description for smaller distances. It is evident that adjusting $d_0$ and $\epsilon$ will not lead to a satisfactory description of the data points over the full range of distances due to the leveling off at small polymer-graphene separations. This indicates another physical mechanism that becomes important for smaller distances. 
It has been noted earlier,\cite{Sau2008} that the simple model given in Eq.~(\ref{eq:image1}) based on a point-like charge does not take into account the polarizability of the adsorbate, an effect which should be more pronounced at small distances. To estimate the magnitude and direction of such a polarization effect, we have extended the standard image potential model as to to include the polarizability of the adsorbate system in a simplified manner (see Appendix~\ref{app:image} for details). To this end, we consider that the image charge $q'$ induces a dipole moment in the adsorbate system which in turn gives rise to a dipole field acting on the physical charge. This model (orange line) qualitatively follows the computed $G_0W_0$ values. It agrees with the standard model at large $d$ but deviates from it at small $d$ due to a correction of opposite sign whose leading $1/d^3$-term is proportional to the polarizability of the adsorbate. Thus, the level-renormalizations level off at small $d$ in accordance with the $G_0W_0$. While our extended model provides physical insight as to why the standard electro-static model starts to break down, we emphasize that it is still too crude to capture details such as the shape of the adsorbate or the anisotropy of its polarizability. Thus, it may serve as a tool for qualitative understanding but not for a quantitative predictions.
Moreover, there are also other effects which may cause the standard image charge model of Eq.~(\ref{eq:image1}) to fail for small adsorption distances.
For instance, for polymer-graphene distances smaller than 4.5~{\AA}, the added electron (hole) into the $\pi_1^*$ ($\pi_1$) state at $\Gamma$ starts to extend beyond the position of the fictitious image plane which may also lead to a deviation from the $1/d$ behavior. Such small hybridization effects between polymer and graphene are exemplified by the minor, but clearly visible, dependence of the HOMO and LUMO LDA-energies (left panel of Fig.~\ref{classical-model-f}).
Finally, to obtain the full electronic band structure, one needs to determine the individual self-energies for each k point and band, which 
goes far beyond the capability of such model.

\section{Summary}

In this paper, we have investigated the electronic properties of the polymer poly({\emph{para}-phenylene) (PPP) adsorbed on graphene by means of \emph{ab initio} electronic-structure calculations. Analysis of the charge density shows the formation of weak interface dipoles, but no net charge transfer between PPP and graphene. At the level of density-functional theory within the local-density approximation, we find that the adsorption of PPP on graphene does not alter the PPP band structure compared to an isolated PPP chain. However, by incorporating many-body effects within the $G_0W_0$ approximation, we obtain a considerable reduction of its HOMO-LUMO gap upon adsorption even for large distances from graphene where the wavefunction overlap between graphene and PPP is negligible. We find that a classical image-potential model in its standard form describes the $G_0W_0$ HOMO and LUMO energies of PPP only for fairly large distances, while we observe some deviations from the expected $1/d$ dependence for polymer-graphene separations smaller than 4.5~{\AA} down to the equilibrium van-der-Waals bond distance of 3.25~{\AA}. 
By incorporating the polarizability of the adsorbate into the model, we are able to qualitatively improve it also for shorter polymer-graphene distances, thereby identifying the most-likely physical mechanisms for the substrate-induced level renormalizations of the polymer HOMO and LUMO close to equilibrium bonding distances. However, for quantitative description $G_0W_0$ calculations are necessary to properly predict the electronic structure of the adsorbates close to surfaces.

\section*{Acknowledgement}
We acknowledge financial support from the Austrian Science Fund (FWF), projects S9714 and P23190-N16.
P.~A.~acknowledges financial support from the Ministry of Science of Iran.

\appendix
\section{Convergence tests for monolayer graphene}

\begin{figure}
\includegraphics*[width=0.8\columnwidth]{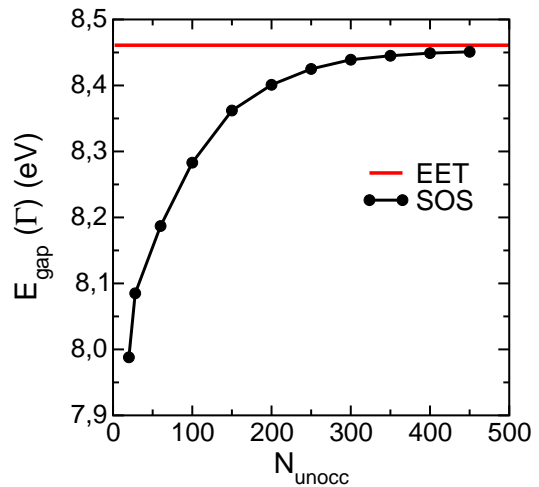}
\caption{\label{sos-gap}
Comparison of the $G_0W_0$ gap of graphene at the $\Gamma$ point using the conventional sum-over-empty-states approach  (circles and solid black line) and the energy effective technique (red line).}
\end{figure}

As discussed in Section \ref{sec:comp}, $G_0W_0$ calculations typically involve time-consuming summations over empty states. Here, we demonstrate for the case of graphene the convergence of the QP energies with respect to the number of empty states and compare with results from the energy effective technique in which all summations are restricted to occupied states only. This is shown in Fig.~\ref{sos-gap} for the QP energy differences between the highest occupied and the lowest unoccupied band at the $\Gamma$ point. Note that for this convergence test, we have used a moderately dense sampling of only $6\times6\times1$ $k$-points. One can see that the QP energy differences obtained by both methods converge to the same value. In view of this very good agreement and the computational efficiency of the EET, we apply this technique in all $G_0W_0$ calculations presented in this paper. 

\begin{figure}
\includegraphics*[width=\columnwidth]{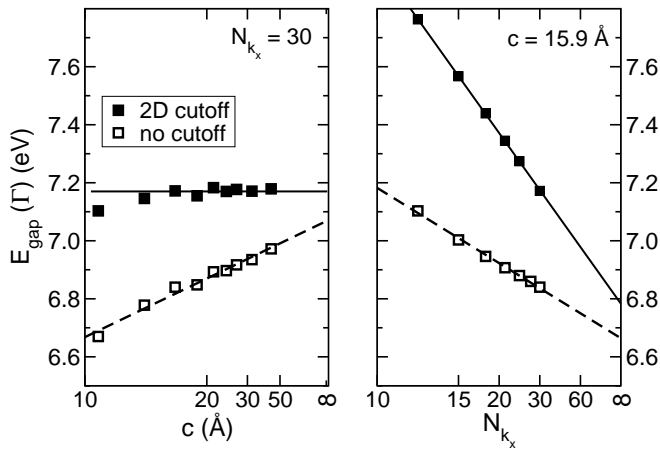}
\caption{\label{qp-gap-gr}
The convergence of the $G_0W_0$ band gap at $\Gamma$ point for monolayer graphene when using a cut-off version of the Coulomb potential (filled symbols) compared to the plain Coulomb potential (open symbols). The left panel shows the convergence as a function of vacuum-layer thickness $c$ for a fixed $k$-point sampling of $30 \times 30 \times 1$ while the right panel displays the convergence as a function of $k$-points ($N_{k_x} \times N_{k_x} \times 1$ meshes) for a fixed vacuum layer thickness of $c=15.9$ \AA.}
\end{figure}

As outlined in Sec.~\ref{sec:comp}, the nonlocal nature of the self-energy leads to a slow convergence of results as a function vacuum layer size in slab geometries
which can be mediated by using  a truncated Coulomb potential. The left panel of Fig.~\ref{qp-gap-gr} shows the convergence of the 
$G_0W_0$ gap at the $\Gamma$ point of hexagonal graphene. The open squares display results for the unmodified Coulomb potential while the filled squares are values obatined by using a cut-off Coulomb potential for slab geometries according to Ismail-Beigi.\cite{Ismail-Beigi2006}
Clearly, the quasi-particle gap is converged at a graphene layer separation of about $c=20$~{\AA} when using the modified Coulomb potential, while it approaches a converged value only via a $1/c$ dependence in the unaltered Coulomb potential (note the reciprocal abscissa in Fig.~\ref{qp-gap-gr}). For instance, at 20~\AA, the error would amount to about 0.2 eV this particular quasi-particle energy difference. It should be noted, however, that the $k$-point convergence worsens when utilizing a cut-off Coulomb potential. This is visualized in the right panel of Fig.~\ref{qp-gap-gr}, where we plot the $G_0W_0$ gap of graphene at $\Gamma$ as a function of $N_{k_x}$ which defines the $k$ mesh ($N_{k_x} \times N_{k_x} \times 1$ points). From the analysis of Fig.~\ref{qp-gap-gr} we conclude that a vacuum layer thickness of $20$~{\AA} and a $k$-mesh of $36 \times 36 \times 1$ should be sufficient to converge quasi-particle energies to within 0.1 eV when employing a cut-off Coulomb potential. For the supercell used in the calculation of the polymer and the combined polymer-graphene system, this $k$-mesh translates into a mesh of $5 \times 17 \times 1$ which we have utilized throughout the paper for all computations involving the supercell.

\section{Image-potential model}
\label{app:image}
The classical image-potential model leading to Eq.~(\ref{eq:image1}) assumes a \emph{point charge} in front of a semi-infinite ($z < 0$), polarizable medium described by the dielectric constant $\epsilon$. A charge $q$ which resides a distance $z=d$ above the dielectric polarizes the dielectric, an effect which is accounted for by an image charge $q'=-q(\epsilon-1)/(\epsilon+1)$ appearing at $z=-d$. When integrating the attractive force between the charge and its image $F(d)=\frac{q q'}{(2d)^2}$ from $d$ to infinity, the energy correction given in Eq.~(\ref{eq:image1}) is obtained. Note that we use atomic units throughout, \emph{i.e.}, we set $e^2/4 \pi \epsilon_0 \rightarrow 1$.

\begin{figure}
\includegraphics*[width=0.8\columnwidth]{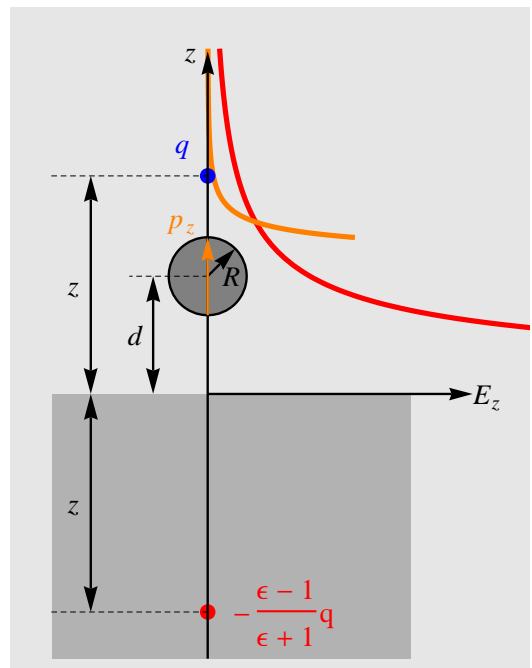}
\caption{\label{imagemodel}
Schematics showing the image-potential model. A charge $q$ at position $z$ above a dielectric with dielectric constant $\epsilon$ filling the half space $z<0$ induces an image charge $q'=-\frac{\epsilon-1}{\epsilon+1}q$ which, in turn, induces a dipole with dipole moment $p_z$ at the location $d$ of the adsorbate which  is modeled by a homogeneous polarizable sphere of radius $R$. The electric fields arising from the image charge (red line) as well as from the image-charge-induced dipole (orange line) are illustrated for $d=4$~{\AA}, $\alpha=15$~a.u. and $R=2.5$~a.u.}
\end{figure}

Let us now extend this simple point-charge description by allowing the adsorbate system to be polarizable. As indicated in Fig.~\ref{imagemodel}, we imagine an adsorbed atom at a fixed distance $d$ above the semi-infinite dielectric. When adding a charge $q$ at distance $z$, the induced charge $q'$ at $-z$ will now \emph{polarize} the atom thereby inducing a dipole moment $p_z$ of strength
\begin{equation}
p_z = \alpha E_z.
\end{equation}
Here, $\alpha$ denotes the polarizability of the adsorbate, and $E_z$ is the $z$-component of the electric field at the location of the physical charge due to the image charge $q'$, thus
\begin{equation}
E_z=\frac{q'}{(z+d)^2}.
\end{equation}
This induced dipole gives rise to an electric field at $z$ which exerts a force on the charge $q$ in addition of the image force due to $q'$. Combining the above expressions one finds for the $z$-component of the total force
\begin{equation}
F_z = q q' \left[ \frac{1}{ (2z)^2} + \frac{2 \alpha}{(z+d)^2 |z-d|^3} \right].
\end{equation}
Here, the first term (red line in Fig.~\ref{imagemodel}) describes the attraction of the charge $q$ by the image charge $q'$, while the second term (orange line), which is proportional to the adsorbate's polarizability $\alpha$, arises due to the image-charge-induced dipole field. When integrating the force from $d$ to $\infty$, the first term gives Eq.~(\ref{eq:image1}) while the second term produces an expression whose leading term is proportional to $1/d^3$. Two things should be noted at this stage. First, the induced dipole at $z=d$ in turn results in an image dipole at $z=-d$ which, when taken into account in self-consistent manner, exerts a force on the charge $q$ at $z$. However, this effect which is proportional to $\alpha^2$ modifies the final result only at leading order of $1/d^6$, \emph{i.e.}, at very short distances due to the faster decay of the dipole compared to the monopole field. 
As a second note, the dipole field proportional to $1/|z-d|^3$ would lead to a diverging force if the charge $q$ approaches the position of the atom $z=d$. In order to circumvent this problem arising from a too simple model, we represent the polarizable adsorbate by a homogeneous, dielectric sphere of radius $R$ exhibiting the polarizability $\alpha$. Then, the field inside the sphere is behaves regularly at $z\rightarrow d$ and the force may be readily integrated yielding  the energy renormalization $V_{im}$ as a function of adsorbate-substrate separation $d$
\begin{equation}
V_{im}(d) = q q' \left[\frac{1}{4 d} - \frac{3 \alpha}{8 R d^3} + \mathcal{O}(\frac{1}{d^4}) \right]
\end{equation} 
The first term gives again the standard point-charge-result, while the remaining terms starting with 
$1/d^3$ are due to polarization effects of the adsorbate and consequently contain the polarizability $\alpha$ and the size $R$ of the adsorbate system. 
When further allowing for a shift of the image plane to the position $d_0$, values of $\alpha=15$~a.u. and $R=2.5$~a.u.~result in the orange curve shown in the right panel of Fig.~\ref{classical-model-f}. 
We note that the atomic polarizability of an isolated carbon atom is about 11~a.u.,\cite{Schwerdtfeger2006} and emphasize that these numerical values merely serve to underpin the main physical effect rather than to create quantitative predictions which would require a more refined electro-static model which takes into account the actual shape, the inhomogeneity as well as the anisotropy of the adsorbate system.

%%%%%%%%%%%%%%%%%%%%%%%%%%%%%%%%%%%%%%%%%%%%%%%%%%%%%%%%%%%%%%%%%%%%%%%%%%%%%%%%%%%%%%%%%%%
%

\end{document}